\documentclass{elsart1p}
\usepackage{graphicx}
\usepackage{amssymb}
\begin{document}
\begin{frontmatter}
%
%
%
\title{Study of Interaction of the Heavy Quarks with Nuclear Matter in Cu+Cu at $\sqrt{s_{NN}}$=200 GeV}
%
%
\author{Miroslav Kr\r us (for the STAR Collaboration)}
\address{Czech Technical University in Prague, Faculty of Nuclear Sciences and Physical Engineering, B\v rehov\' a~7, 115 19 Prague, Czech Republic}
\address{Nuclear Physics Institute of the Academy of Sciences of the Czech Republic v. v. i., 250 68 \v Re\v z}
\begin{abstract}
In this paper we present the study of the azimuthal correlation function of non-photonic electrons with low-p$_\mathrm{T}$ hadrons produced in Cu+Cu collision at $\sqrt{s_{NN}}$~=~200~GeV measured by STAR experiment at RHIC.  Possible modification of the awayside peak is observed.
\end{abstract}
\end{frontmatter}
%
\section{Introduction}
\label{introduction}
Recent STAR and PHENIX experimental results \cite{abelev, horner, adler} have shown that light partons with
high transverse momentum lose a significant amount of energy \cite{energ1, energ2} when
traversing the dense nuclear medium created in central Au+Au collisions.
This leads to suppression of high-p$_\mathrm{T}$ hadron yields and the
disappearance of the away-side peak in azimuthal hadron-hadron
correlation function. The investigation of the azimuthal correlation function of high-p$_\mathrm{T}$ hadrons with medium- or low-p$_\mathrm{T}$ associated particle have shown on the away side ($\Delta \phi = \pi$) a broad double-peak structure. There is intensive discussion of the possible explanation of this observation including Mach cone scenario \cite{machcone}, gluon Cherenkov radiation \cite{cherenkov}, or jet deflection \cite{deflect}.

An interesting question is whether similar effect is present in a case of heavy quarks passing the medium. Heavy quarks are primarily produced during early stages of a nuclear collision, then they interact with the medium, and they can be used as a probe of the space-time evolution of the medium arising from a heavy ion collision. Because of their large masses, they thermalize later and their energy loss is expected to be influenced by dead-cone effect and elastic energy loss.

\section{Electron identification}
\label{identification}
The direct identification of heavy quark mesons (D, B) is difficult with current detectors and RHIC luminosities. Therefore most previous studies used non-photonic electrons consisting of electrons from semileptonic decays of D and B. The identification uses semileptonic decays of open heavy flavor mesons (e.g. $D^0 \rightarrow e^{+}K^{-}\nu_e$) over broad p$_\mathrm{T}$-range. These non-photonic electrons keep well direction of the mother heavy meson when electron has p$_\mathrm{T} > $ 3 GeV/c \cite{gang}.

For results presented here the data from Cu+Cu at $\sqrt{s_{NN}}$ = 200 GeV measured in 2005 by STAR are used. The analysis steps of the electron identification \cite{jaro} has been reported in details previously. In general, electrons are identified by the combination of the TPC \cite{tpc} (Time Projection Chamber) ionization energy loss, the deposited energy in the BEMC \cite{barrel}(Barrel Electromagnetic Calorimeter), and the electromagnetic shower profile in the SMD \cite{barrel} (Shower Maximum Detector).

The major difficulty in the electron analysis is the fact that there are many sources of the electrons other than semileptonic decays of heavy mesons, for instance from photon conversions in the detector material between the interaction point and the TPC and $\pi^0$ and/or $\eta$ Dalitz decays. These photonic electrons are identified by the invariant mass distribution of electron-positron pairs. The photonic electron yield is given by the difference between the opposite- and same-sign distribution below the invariant mass cut (typically $\pi^0$ mass). The same-sign distribution is due to combinations of random pairs. The purity of electron sample is above 99\% \cite{gang}.

\section{Non-photonic electron-hadron correlations}
\label{correlations}

The study of the azimuthal non-photonic electron-hadron correlations (trigger electron with $3<\mathrm{p}_{\mathrm{T}}^{\mathrm{trig}}<6$ GeV/c and associated charged hadron with 0.15$< \mathrm{p}_{\mathrm{T}}^{\mathrm{assoc}}<0.5$ GeV/c) uses the semi-inclusive electron sample \cite{gang}, when electrons with the opposite-sign partner with an invariant mass cut are excluded from the inclusive electron sample. The non-photonic electron-hadron correlations ($\Delta \Phi_{NP}$) are obtained by the formula
\begin{displaymath}
 \Delta \Phi_{NP}=\Delta \Phi_{SI}+\Delta \Phi_{SS}-\left( \frac{1}{\varepsilon}-1 \right) \left( \Delta \Phi_{OS}-\Delta \Phi_{SS} \right),
\end{displaymath}
where each term represents correlation functions of the individual electron samples (SI - semi-inclusive, SS - same-sign, and OS - opposite-sign) with charged hadrons and $\varepsilon$ is the efficiency of the photonic electron reconstruction estimated by embedding simulated data into real events.

\begin{figure}
\begin{center}
 \includegraphics[width = 12 cm]{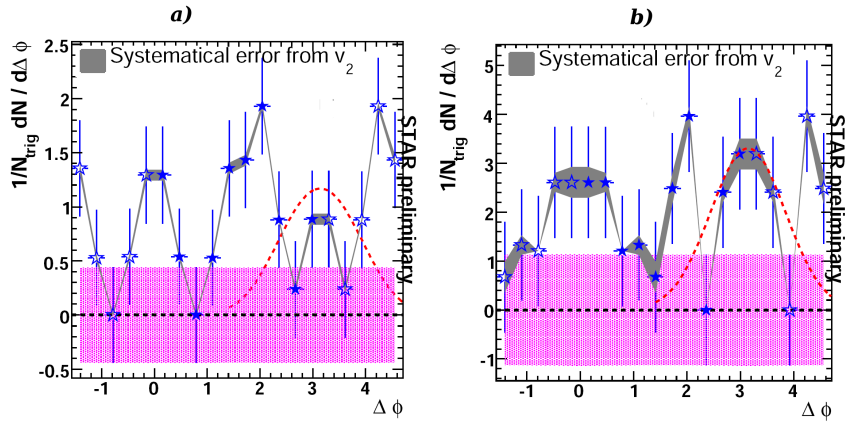}

\caption{Non-photonic electron-hadron correlations. Panel a) shows correlation in Cu+Cu and panel b) in Au+Au collisions at 200 GeV after $v_2$ subtraction ($v_2$ = 0.05). The error bars are statistical and the error band represents ZYAM systematical uncertainty. The dashed curve is the PYTHIA prediction of the away side peak.}
\label{pict}
\end{center}

\end{figure}

Despite large statistical errors, one can see clear correlation structure of non-photonic electrons with hadrons in central (centrality 0 - 20\%) Cu+Cu collisions (Fig. \ref{pict}). On the nearside ($\Delta \phi$ = 0), the single peak  represents the heavy quark fragmentation. On the awayside ($\Delta \phi$ = $\pi$), the correlation function shows a broad or possible double-hump structure. The elliptic flow (v$_{2}$) background was subtracted with the use of the ZYAM \cite{zyam} method. Similar pattern has been seen in central Au+Au collisions (centrality 0 - 20\%) \cite{gang} (Fig. \ref{pict}).

\section{Summary}
\label{sumary}
The broad modification of the awayside peak in both Cu+Cu and Au+Au central collisions is similar to the di-hadron correlations in Au+Au, and probably indicates interaction of heavy quarks with the dense medium.

\section*{Acknowledgments}
\label{acknow}
This work was supported in part by the IRP AV0Z10480505, by GACR grant 202/07/0079 and by grants LC07048 and LA09013
of the Ministry of Education, Youth and Sports of the Czech Republic.
%
%
%

%
\end{document}